\documentclass[aps,pra,reprint,showpacs,groupedaddress]{revtex4-1}

\usepackage{graphicx}
\usepackage{amsmath,amssymb}
\begin{document}

\title{Laser Phase and Frequency Stabilization using Atomic Coherence}
\author{Yoshio Torii}
\email[]{ytorii@phys.c.u-tokyo.ac.jp}
\author{Hideyasu Tashiro}
\author{Nozomi Ohtsubo}
\author{Takatoshi Aoki}
\affiliation{Institute of Physics, The University of Tokyo, 3-8-1 Komaba, Meguro-ku, Tokyo 153-8902, Japan}
\date{\today}

\begin{abstract}
We present a novel and simple method of stabilizing the laser phase and frequency by polarization spectroscopy of an atomic vapor. In analogy to the Pound-Drever-Hall method, which uses a cavity as a memory of the laser phase, this method uses atomic coherence (dipole oscillations) as a phase memory of the transmitting laser field. A preliminary experiment using a distributed feedback laser diode and a rubidium vapor cell demonstrates a shot-noise-limited laser linewidth reduction (from 2 MHz to 20 kHz). This method would improve the performance of gas-cell-based optical atomic clocks and magnetometers and facilitate laser-cooling experiments using narrow transitions.
\end{abstract}
\maketitle

\section{Introduction}
The reduction of the laser linewidth has always been an essential and critical issue in atomic, molecular, and optical physics since the first realization of lasers in 1960. In laser-cooling experiments, for example, the laser linewidth should be narrower than the natural width of the cooling transition of atoms (ranging from kHz to MHz) for Doppler cooling to work \cite{Metcalf}. The importance of narrow-linewidth lasers is more prominent in the field of optical atomic clocks, where an extremely narrow ($\sim$mHz) optical transition of atoms \cite{Takamoto} or single ions \cite{Rosenband} is probed by a laser with a sub-Hertz linewidth \cite{Young, Stoehr}.

The Pound-Drever-Hall (PDH) method, invented in the early 1980s \cite{Drever}, has become the standard method for laser linewidth reduction. The key feature of the PDH method is its use of the cavity not only as a frequency reference but also as a {\it time-averaged phase memory} of the incident laser field, which leads to a fast response to the laser phase fluctuations on a time scale shorter than the cavity response time, enabling efficient laser linewidth reduction \cite{Drever, Black}. We note that the H\"{a}nsch-Couillaud laser stabilization scheme \cite{Hansch}, in which the change of the reflected light polarization is monitored, also benefits the phase-storing property of the cavity.

In this paper, we propose an alternative method of laser phase and frequency stabilization, in which {\it atomic coherence} (dipole oscillations of atoms) is used as a phase memory of the laser field transmitting through an atomic vapor. This method has an apparent advantage of its immunity to mechanical noise from environment. Moreover, unlike cavities made of glass with heat expansion, the resonance frequency of atoms, to which the laser is locked, is independent of the temperature of the atomic vapor in a Doppler-free configuration. These features would drastically simplify the experiments which have relied on the PDH method using an ultralow heat expansion (ULE) cavity for laser linewidth reduction \cite{Young, Stoehr, Alnis}. We preformed a preliminary experiment using a distributed feedback (DFB) laser diode and a rubidium vapor cell, and observed a significant linewidth reduction (from 2 MHz to 20 kHz), proving the effectiveness of the method. 

\section{Principle of the method}
\begin{figure}[bbbb]
\begin{center}
\includegraphics[width=85mm]{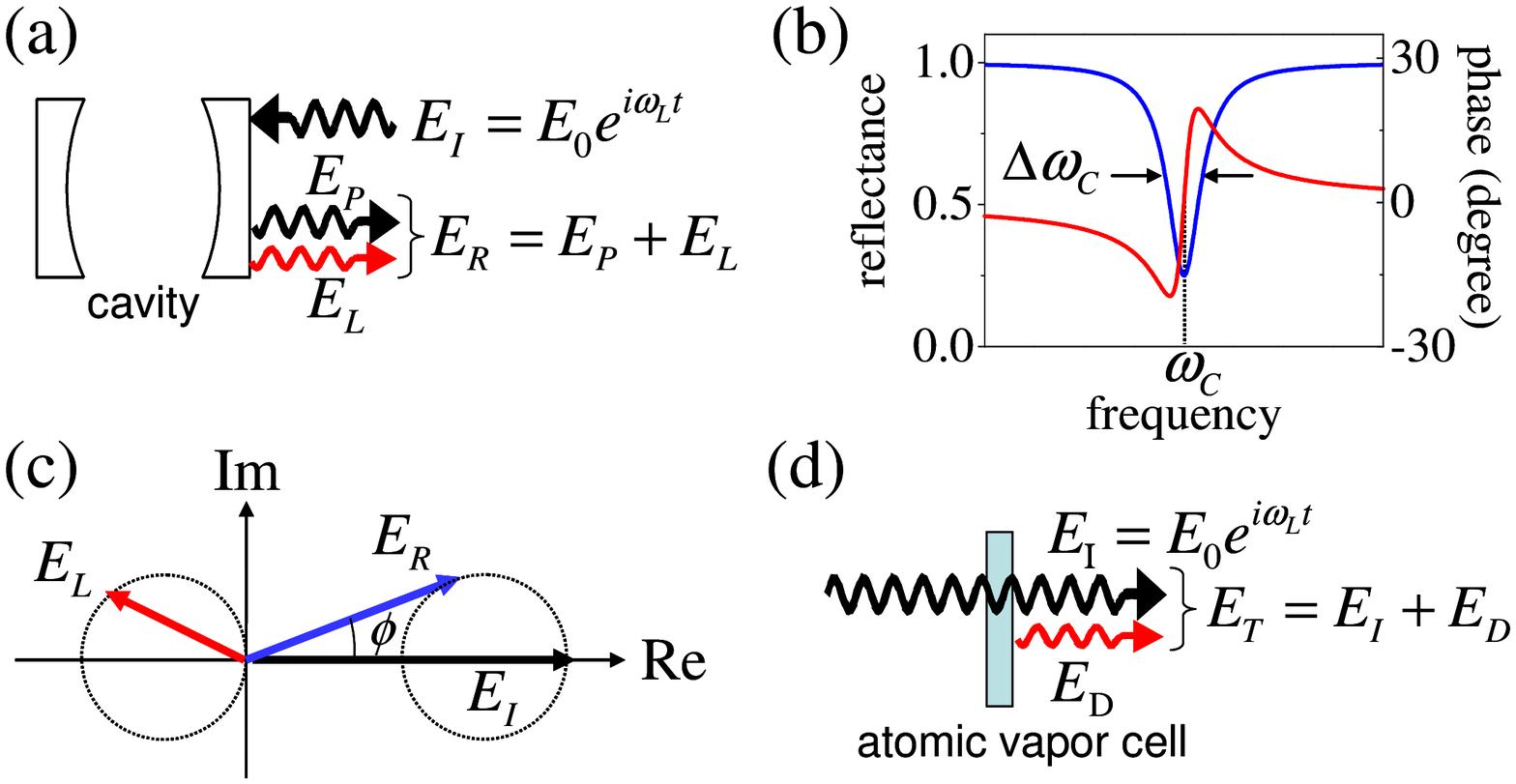}
\end{center}
\caption{(Color online) (a) Relevant fields in the PDH laser stabilization.  (b) Reflectance (blue line) and phase (red line) of the reflected field. (c) Relation between relevant fields in the complex plane. (d) Relevant fields in laser stabilization using an atomic vapor. }
\label{fig1}
\end{figure}

To show the analogy between the PDH method and the proposed method, let us first review the principle of PDH phase and frequency stabilization in detail. Suppose a laser beam with frequency $\omega _{L}$ is incident to a cavity with resonance frequency $\omega_{C}$ and resonance width (FWHM) $\Delta {{\omega }_{C}}$ as in Fig. \ref{fig1}(a). 
We write the complex amplitude of the incident field, measured at a point just outside the cavity, as $E_{I}= E_{0}\exp (i{{\omega }_{L}}t)$. The reflected field $E_{R}$ is the sum of the promptly reflected field $E_{P}$ off the input mirror and the leakage field $E_{L}$ from the cavity. In the case of a high finesse cavity, $E_{P}\approx E_{I}$ (we assume there is no phase change upon reflection for simplicity) and the steady-state complex amplitude of the reflected field is then expressed as \cite{Hansch}
\begin{equation}
	E_{R}= E_{P} +E_{L} = \left( 1-\frac{1-{{R}_{1}}}{1-R}\frac{1-ix}{1+{{x}^{2}} }\right) E_{I},
\label{reflectance}
\end{equation}
where $R_{1}$ is the intensity reflectivity of the input mirror, $R$ is the amplitude ratio between successive roundtrips in the cavity, and $x=(\omega_{L}-\omega_{C})/(\Delta \omega _{C}/2)$ is a normalized laser frequency detuning from the cavity resonance. In the following, we assume $R<R_{1}$ (undercoupled cavity) where the reflected light remains even at the center of the resonance as depicted in Fig. \ref{fig1}(b). 

Figure \ref{fig1}(c) shows the relation between ${{E}_{I}}$, ${{E}_{R}}$, and ${{E}_{L}}$ in the rotating complex plane where ${{E}_{I}}$ always lies on the positive side of the real axis. The reflected field ${{E}_{R}}$, given by Eq. (\ref{reflectance}), traces out a circle clockwise as the laser frequency $\omega _{L}$ increases, and the resultant phase shift of the reflected field (with respect to the incident field) shows a dispersive curve centered at the cavity resonance as depicted in Fig. \ref{fig1}(b) \cite{Black}. In the PDH method, this phase shift is converted to a dc electric signal by frequency modulation (FM) spectroscopy \cite{Bjorklund}, and used as a frequency discrimination signal.

If the phase of the incident laser makes a positive (negative) jump on a time scale shorter than the cavity response time ${{\tau }_{C}}=1/\Delta \omega {}_{C}$, the field inside the cavity can not follow this phase jump and the leakage field relatively shifts clockwise (counterclockwise) in the rotating complex plane [Fig. \ref{fig1}(c) shows a case of positive phase jump], resulting in an error signal with the same sign as a positive (negative) frequency shift. This phase-sensitive response at short times $<{{\tau }_{C}}$ is the heart of the PDH laser phase stabilization.

The phase-sensitive response of the PDH frequency discriminator can be explained more formally by its transfer function $G({{\omega }_{M}})={{G}_{0}}/(1+2i{{\omega }_{M}}/{{\omega }_{C}})$, where ${{\omega }_{M}}$ is the {\it frequency of the laser-frequency modulation} and ${{G}_{0}}$ is the gain in the passband \cite{Nagourney}. The characteristic single-poll roll-off (6 dB/octave) and 90${}^\circ$ phase lag above the cutoff frequency ($\Delta \omega_{C}/2$) ensure the prompt response of the discriminator to a sudden phase jump \cite{Zhu}.

Now we consider the case where the role of the cavity is replaced by an atomic vapor. Suppose a laser beam with complex amplitude $E_{I}= E_{0}\exp (i{{\omega }_{L}}t)$ is incident onto an atomic vapor cell of length $l$ containing two-level atoms with resonance frequency ${{\omega }_{A}}$ and resonance width (FWHM) $\Gamma $. We assume here that the complex electric susceptibility of the atomic vapor has a textbook form of nonsaturated two-level atoms: $\chi ={\chi }'+i{\chi }''=-{{{\chi }''}_{\max }}(x+i)/(1+{{x}^{2}})$, where $x=({{\omega }_{L}}-{{\omega }_{A}})/(\Gamma /2)$ is a normalized laser frequency detuning and ${{{\chi }''}_{\max }}$ is the maximum of the imaginary part of $\chi$. If the vapor cell is optically thin and the attenuation of the incident laser field is little, the transmitted field $E_{T}$ is, as depicted in Fig.\ref{fig1}(d), approximately given by the sum of the incident field $E_{I}$, which travels as if there were no atoms in the cell, and the field $E_{D}$ produced by incident-field-induced dipole oscillations of atoms \cite{Feynman}. The steady-state complex amplitude of the transmitted field is then expressed as
\begin{equation}
E_{T}= E_{I}+ E_{D}=\left( 1-\frac{\alpha l}{2 }\frac{1-ix}{1+{{x}^{2}} } \right) E_{I},
\label{dipole}
\end{equation}
where $\alpha =k{{\chi }''}_{\max }$ is the maximum absorption coefficient of the atomic vapor and $k$ is the wave number of the incident laser field. One can find that the transmitted field $E_{T}$ has exactly the same form as the cavity reflection field $E_{R}$ given by Eq.(\ref{reflectance}) and the dipole field $E_{D}$ plays the same role as the cavity leakage field $E_{L}$. This mathematical equivalence between the cavity and the thin atomic vapor, combined with the fact that the atomic dipole oscillations cannot follow a sudden phase change of the incident field for shorter times than the atomic coherence time ${{\tau }_{A}}=1/\Gamma $, naturally leads us to a conclusion that laser phase and frequency stabilization using an optically thin atomic vapor would work.

The above conclusion can be extended to an atomic vapor cell whose optical density is $\sim$1. As long as we consider the time scale shorter than the atomic coherence time, each atomic dipole at a different part of the cell, after a sudden phase change of the incident field, still continues oscillating with the previous phase and contributes independently to the shift of the complex amplitude of the transmitted field, thus enabling a laser phase locking to the atomic dipole oscillations.

The phase shift of the transmitted field can be converted to an electric signal by FM spectroscopy as in the PDH method. Alternatively, the phase shift can be converted to a polarization rotation of the transmitted laser beam by polarization spectroscopy \cite{Wieman}. If the atoms are prepared such that only the $\sigma ^{+}(\sigma ^{-})$ components of the linearly polarized incident laser beam interacts with the atoms and the $\sigma ^{-}(\sigma ^{+})$ component just serves as a phase reference, the polarization rotation angle is given by $\Delta \theta =\phi /2$, where $\phi $ is the phase shift of the $\sigma ^{+}(\sigma ^{-})$ component of the transmitted laser beam.  A balanced polarimeter can convert this polarization rotation to a photocurrent signal $\Delta I$ with a relation $\Delta I=2{{I}_{T}}\Delta \theta ={{I}_{T}} \phi $, where ${{I}_{T}}$ is the photocurrent signal corresponding to the total transmitted beam power \cite{Yoshikawa}. This phase-to-photocurrent conversion by polarization spectroscopy was utilized in the experiment described below.

The transfer function of a frequency discriminator based on polarization spectroscopy can be derived using Eq. (\ref{dipole}) on the assumption that the intensity of the incident laser beam is well below the saturation intensity and the atomic vapor is optically thin. If we suppose that the instantaneous frequency of the incident laser beam has a time dependence of $\omega(t) = {\omega}_{A}+\Delta {{\omega }_{0}}\cos {{\omega }_{M}}t$, where $\Delta {\omega }_{0}<\Gamma$ is the maximum frequency deviation and ${\omega }_{M}$ is the modulation frequency, the resultant time dependence of the photocurrent signal at the balanced polarimeter is calculated to be  
\begin{equation}
\label{transfer}
\Delta I={{I}_{T}}\frac{\alpha l\Delta {{\omega }_{0}}}{(1-\alpha l/2)\Gamma }\frac{x_{M}^{{}}\sin {{\omega }_{M}}t+\cos {{\omega }_{M}}t}{1+x_{M}^{2}},
\end{equation}
where ${{x}_{M}}={{\omega }_{M}}/(\Gamma /2)$ is a normalized modulation frequency (see Appendix for a detailed derivation). Equation (\ref{transfer}) shows that the transfer function of the frequency discriminator is $G({{\omega }_{M}})={{G}_{0}}/(1+2i{{\omega }_{M}}/\Gamma )$, where ${{G}_{0}}={{I}_{T}}\alpha l/(1-\alpha l/2)\Gamma$. This transfer function has exactly the same form as that for a PDH frequency discriminator if we replace $\Gamma$ with $\Delta \omega_{C}$, supporting the above conclusion that laser phase and frequency stabilization using atomic dipole oscillations would work. 

We note that the conversion of a phase jump in the incident field into a shift of the complex amplitude of the transmitted field by an atomic vapor discussed above has the same origin as the phenomena known as frequency (FM) noise to amplitude (AM) noise conversion \cite{Camparo1999}, or FM noise to optical rotation (OR) noise conversion \cite{Kitching2009}; the latter exactly corresponds to the case we have considered here. For gas-cell-based optical atomic clocks \cite{Knappe2004} or magnetometers \cite{budker2007}, FM-to-AM or FM-to-OR noise conversion is undesirable because it degrades the signal to noise ratio in the transmitted light and limits the ultimate performance of the device. On the other hand, for our method to work, FM-to-OR noise conversion is essential and inevitable.

\section{Experimental setup}
\begin{figure}
\begin{center}
\includegraphics[width=70mm]{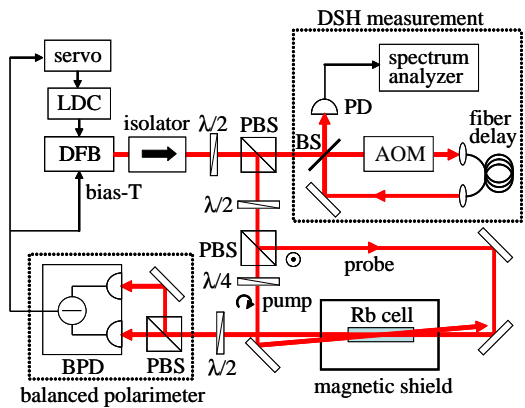}
\end{center}
\caption{(Color online) Experimental setup. LDC, laser diode current controller; BS, 50:50 beamsplitter; PD, photodetector; BPD, balanced photo detector; $\lambda /2$, half-wave plate; $\lambda /4$, quarter-wave plate.}
\label{fig2}
\end{figure}

We performed a preliminary experiment to verify the validity of the proposed method. The experimental arrangement is shown in Fig. \ref{fig2}. 
To emphasize the generality and simplicity of the method, we equipped the system with commercial electronic devices except a homemade low-bandwidth servo circuit. As a light source to be phase and frequency stabilized, we chose a DFB laser diode operating at 780 nm (Eagleyard EYP-DFB-0780-00080) with an intrinsic linewidth of about 2 MHz \cite{Kraft}. The DFB laser was installed in a commercial mount (Thorlabs TCLDM9) featuring a bias-T current modulation input with a bandwidth of 0.2 to 500 MHz, and driven by a low-noise laser diode current controller (Thorlabs LDC201ULN). Most of the power ($\sim$20 mW) in the beam from the DFB laser was used for laser linewidth measurement as described below, and a small portion ($\sim$2 mW) of the beam was used for laser phase and frequency stabilization by polarization spectroscopy of the Rb D2 line (natural width $\Gamma = 2\pi \times 6$ MHz) \cite{Do}.

The optical setup was basically the same as the previous works which demonstrated  laser frequency stabilization using polarization spectroscopy \cite{Yoshikawa, Pearman}. A circularly polarized pump beam and a linearly polarized probe beam, both of which had almost the same diameters ($\sim$5 mm) and powers ($\sim$1 mW), were sent to a Rb cell at room temperature in a Doppler-free configuration. The Rb vapor cell was 75 mm long and situated in a magnetic shield. A balanced polarimeter consisted of a balanced photo detector (Thorlabs PBD150A) and a polarizing beamsplitter (PBS). The bandwidth and the gain of the balanced photo detector were set to 50 MHz and $10^{4}$ V/A ($\sim $0.5 V/mW for 780 nm), respectively.

Figure 3(a) shows a polarization rotation (PR) spectrum when the laser frequency was scanned around the 5S$_{1/2},F=2\to$ 5P$_{3/2},{F}'=1,2,3$ transitions of $^{85}$Rb. 
\begin{figure}[tttt]
\begin{center}
\includegraphics[width=80mm]{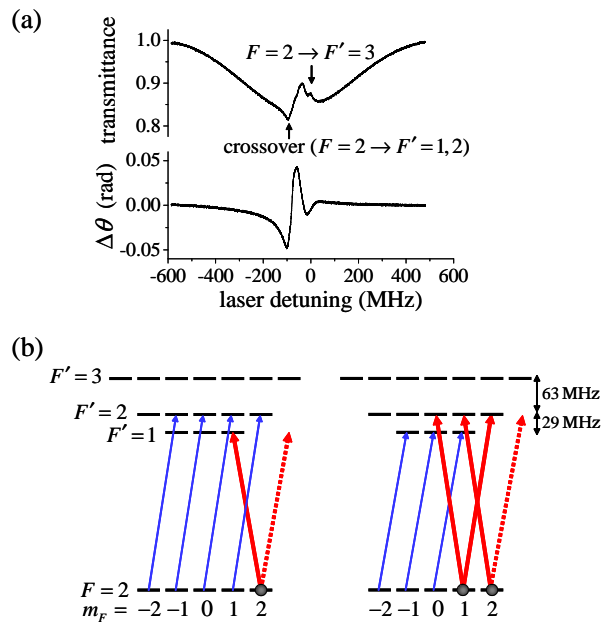}
\end{center}
\caption{(Color online) (a) Saturated-absorption spectrum (upper trace) and polarization rotation spectrum (lower trace) scanned around the 5S$_{1/2},F=2\to$ 5P$_{3/2},{F}'=1, 2, 3$ transitions of $^{85}$Rb. (b) Two pump-probe configurations contributing to the crossover transition. The blue thin arrows indicate the transitions by the $\sigma ^{+}$ pump (spin-polarizing) beam which optically pumps the atom to the ${{m}_{F}}=1$ or ${{m}_{F}}=2$ states. The red thick arrows indicate the transitions by the probe beam. The $\sigma ^{+}$ transitions which are not allowed for the spin-polarized atom are indicated by the red dotted arrows. In both configurations, the $\sigma ^{-}$ component of the probe beam predominantly interacts with the atoms, resulting in circular birefringence for the probe beam.}
\label{fig3}
\end{figure}
A saturated-absorption (SA) spectrum was also taken just by rotating the half-wave plate in front of the balanced polarimeter such that all the power of the probe beam was detected by one of the balanced photo detector inputs. As has been reported in Ref.\cite{Do}, we observed a large dispersion signal in the PR spectrum at the crossover transition between the $F=2\to{F}'=1$ and the $F=2\to{F}'=2$ transitions. This is because the $\sigma ^{-}$ component of the probe beam, as depicted in Fig.3(b), predominantly interacts with the atoms in both of the two pump-probe configurations which contribute to the crossover transition. The $F=2\to{F}'=1$ and $F=2\to{F}'=2$ transitions were not resolved from the crossover transition in the PR and the SA spectra because the intensities of tＳａｔｕｒａｔｅｄhe pump and the probe beams were comparable with the saturation intensity ($I_{s}=3.57$ mW/cm$^{2}$ for the unresolved $F=2\to{F}'=1, 2$ transition) and the width of each transition was power-broadened to $\sim$10 MHz \cite{Note0}. If we assume that the observed maximum absorption of 18\% in the SA spectrum \cite{Note1} was solely due to the absorption of the $\sigma ^{-}$ component of the probe beam, the maximum optical density of the Rb vapor for this component is ${\text{OD}}_{\max }$ = 0.44. The maximum angle of the polarization rotation is then calculated from Eq.(\ref{dipole}) to be $\Delta {{\theta }_{\max }}={{\phi }_{\max }}/2={\text{OD}}_{\max }/8$ = 0.055 rad, which is close to the observed value of 0.050 rad. 

We performed laser phase and frequency stabilization of the DFB laser using a dispersion signal obtained by PR spectroscopy on the above-mentioned crossover transition of $^{85}$Rb. Bias-T fast current feedback was implemented to suppress fast phase fluctuations of the DFB laser. The output signal from the balanced detector was split and fed to the bias-T current modulation input and the servo circuit which gently locked the laser frequency to the zero-crossing point of the dispersion signal through the external modulation input (bandwidth $\sim$100 Hz) of the current controller.

 The linewidth of the phase and frequency stabilized laser was measured by a delayed self-homodyne (DSH) technique \cite{Okoshi}. As depicted in Fig. \ref{fig2}, the laser beam was split into two beams by a 50:50 beamsplitter and one beam (the signal) was multiply sent to an acousto-optic modulator (AOM) driven at 60 MHz and a 300-m single-mode fiber to increase the delay time, and combined with the other beam (the local oscillator) as demonstrated in Ref.\cite{Tsuchida}. The beat signal was detected by a fast photo diode (Thorlabs DET210) with a response time of 1 ns and fed to a spectrum analyzer. The maximum delay time was limited by the attenuation of the signal beam power in the roundtrips. We could observe the beat signal of up to 300 MHz (five round-trips), corresponding to a delay time of ${{\tau }_{d}}$ = 7.2 $\mu$s. The experimental data presented below were all taken with this delay time.

\section{Results and discussion}
\begin{figure}[tb]
\begin{center}
\includegraphics[width=70mm]{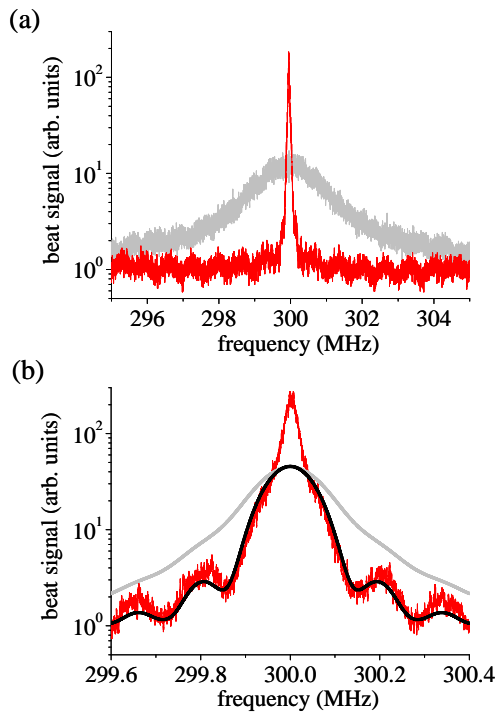}
\end{center}
\caption{(Color online) (a) Power spectra of the beat signals in a delayed self-homodyne measurement using a 780-nm DFB laser. The red and gray lines show the spectra obtained when the bias-T current feedback was on and off, respectively. The spectrum analyzer was set with a resolution bandwidth of 10 kHz and a video bandwidth of 100 Hz. (b) A detailed scan of the central part of the spectrum when the bias-T feedback was on. The resolution bandwidth was reduced to 1 kHz. The red line shows the observed spectrum, while the black and gray lines represent the calculations for laser linewidths (coherence times) of 20 kHz (8 $\mu$s) and 80 kHz (2 $\mu$s), respectively.}
\label{fig4}
\end{figure}

The results of DSH measurements are shown in Fig. \ref{fig4} (a). 
We observed a significant linewidth reduction by the bias-T current feedback. The suppression of the wings of the beat spectrum observed up to $\sim$10 MHz, which is far beyond the cutoff frequency of the discriminator ($\Gamma /2 = 2\pi \times  3$ MHz), indicates the expected fast response of the discriminator to the laser phase fluctuations above the cutoff frequency. Figure \ref{fig4}(b) shows a detailed scan of the central part of the beat spectrum when the bias-T feedback was on. The ripples in the wings of the spectrum, the spacing of which (140 kHz) is given by the inverse of the fiber delay time, is a characteristic behavior observed only when the coherence time of the laser is comparable to or longer than the delay time \cite{Richter}, suggesting that the laser linewidth was narrowed below $1/(2\pi {\tau }_{d})$ = 22 kHz. To be more quantitative, we compared the wings of the spectrum with calculations according to Ref.\cite{Richter} on the assumption that the laser had a Lorentzian lineshape. The observed ripples in the spectrum fit well a calculation with a laser linewidth (FWHM) of 20 kHz [Fig. \ref{fig4}(b)].

The fundamental limit of any laser frequency discriminator is set by the photoelectron shot noise.  For laser frequency $\nu_{L}$ and detected laser power ${{P}_{0}}$, the photon number fluctuation due to the shot noise at the detector for measurement time $\tau $ is $\sqrt{{{{P}_{0}}\tau /h {{\nu }_{L}}}}$, and the limit of frequency discrimination using atoms is easily calculated as 
\begin{equation}
\Delta {{\nu }_{\text{SN}}}=\frac{1}{2}{{\left( \frac{d\theta }{d{\nu }_{L} } \right)}^{-1}} \sqrt{{ \frac{h {{\nu }_{L}}}{{{P}_{0}}\tau } }}=\frac{\Delta \nu }{\text{O} {\text{D}_{\max } }} \sqrt{{ \frac{h {{\nu }_{L}}}{{{P}_{0}}\tau } }},
\label{shotnoise}
\end{equation}
where $d\theta /d{\nu }_{L}=\text{O}{{\text{D}}_{\max }}/2\Delta \nu $ is the slope of the dispersive curve of polarization rotation at resonance,  $\text{O}{{\text{D}}_{\max }}$ is the maximum optical density, and $\Delta \nu$ is the homogeneous linewidth (FWHM) of the atoms. Substituting the experimental values ($d\theta /d{\nu }_{L} = 5\times10^{-3} $ rad/MHz, ${{P}_{0}}=1$ mW, $h{\nu}_{L}=2.5\times 10^{-19}$ J, and $1/\tau=2\pi \times 50$ MHz) into Eq. (\ref{shotnoise}) gives $\Delta {\nu }_{\text{SN}}\sim$30 kHz, suggesting that the observed linewidth of 20 kHz was nearly shot noise limited.

Equation (\ref{shotnoise}) tells us that using a narrower optical transition would result in a narrower shot-noise-limited linewidth as long as the optical density of the atomic vapor is $\sim$1. Taking the 5$^{1}{{\text{S}}_{0}} \to 5{}^{3}{{\text{P}}_{1}}$ intercombination transition of Sr at 689 nm with a natural width of 7.1 kHz, for example, a heat pipe oven operated around 400 ${}^\circ$C was proven to provide an optical density of $\sim$0.1 without optical window degradation \cite{Li}. With the experimental parameters in Ref.\cite{Li} ($\text{O}{{\text{D}}_{\max }}$ = 0.16, $\Delta \nu$ = 130 kHz, and ${{P}_{0}}$ = 50 $\mu$W) and a photodetector bandwidth of 1 MHz, the shot-noise-limited linewidth is calculated to be $\sim$200 Hz. We note that FM spectroscopy is, in general,  preferable to  polarization spectroscopy to achieve a shot-noise-limited linewidth because the latter suffers from 1/f (flicker) noise in the detector and  polarization fluctuation caused by probe laser absorption \cite{Note2}. We are now planning to reduce the linewidth of a 689-nm laser diode below 1 kHz by FM spectroscopy with a heat pipe oven to perform laser cooling of Sr atoms using this narrow transition. 

\section{Conclusion}
We have proposed and demonstrated a novel laser phase and frequency stabilization method which uses atomic coherence and thereby is immune to environmental mechanical noise and thermal drift. A preliminary experiment using Doppler-free polarization spectroscopy of a Rb vapor demonstrated a shot-noise-limited linewidth reduction of a DFB laser. The laser phase and frequency stabilization demonstrated here can be viewed as the suppression of laser frequency (FM) noise to amplitude (AM) noise conversion, or FM noise to optical rotation (OR) noise conversion in a resonant medium. Our method is readily applicable to improve the performance of gas-cell-based optical atomic clocks or magnetometers, which have been suffering from this type of laser-induced noise.  Our method would also facilitate laser-cooling experiments requiring narrow linewidth lasers with little thermal drift.

\begin{acknowledgments}
We would like to thank Y. Yoshikawa, K. Nakayama, M. Takeuchi, and T. Kuga for helpful discussions. This work was supported by Grant-in-Aid for Challenging Exploratory Research (KAKENHI 23656042) from Japan Society for the Promotion of Science (JSPS).
\end{acknowledgments}

\appendix*
\section{Derivation of the transfer function for a polarization-spectroscopy-based frequency discriminator}
\begin{figure}[tttt]
\begin{center}
\includegraphics[width=60mm]{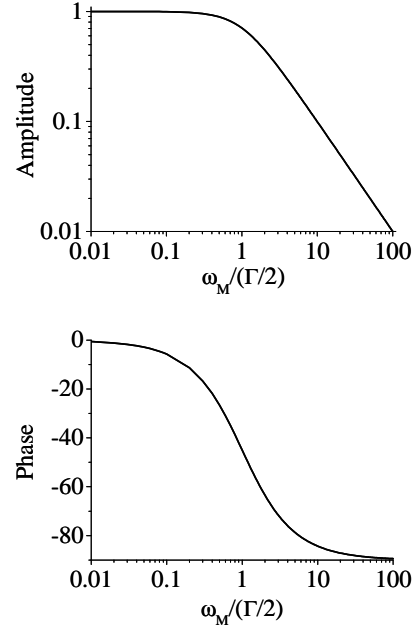}
\end{center}
\caption{Amplitude and phase of the transfer function [Eq.(\ref{transfer2})] for a frequency discriminator based on polarization spectroscopy. The amplitude is normalized by the value at the passband. The horizontal axes represent the modulation frequency normalized by the 3-dB cutoff frequency ($\Gamma/2$).}
\label{figA1}
\end{figure}

Suppose a laser beam with complex amplitude $E_{I}= E_{0}\exp (i{{\omega }_{L}}t)$ is incident onto an atomic vapor cell of length $l$ containing two-level atoms with resonance frequency ${{\omega }_{A}}$ and resonance width (FWHM) $\Gamma $. On the assumption that the intensity of the incident laser beam is well below the saturation intensity and the atomic vapor is optically thin, the steady-state complex amplitude of the transmitted field is expressed as
\begin{equation}
E_{T}=\left( 1-\frac{\alpha l}{2 }\frac{1-ix}{1+{{x}^{2}} } \right) E_{I},
\label{dipole2}
\end{equation}
where $x=({{\omega }_{L}}-{{\omega }_{A}})/(\Gamma /2)$ is a normalized laser frequency detuning and $\alpha$ is the absorption coefficient of the atomic vapor on resonance. To derive the transfer function of a frequency discriminator based on polarization spectroscopy, we first calculate the transmitted field ${{E}_{T}}$ when the incident field  ${{E}_{I}}$ is frequency modulated at ${{\omega }_{M}}$. We suppose here that the (carrier) frequency of the incident laser is tuned to the atomic resonance at $\omega_{A}$. In general, a frequency-modulated laser can be thought of as a phase-modulated laser. The frequency-modulated incident laser field can be expressed as
\begin{equation}
{{E}_{I}}={{E}_{0}}\exp \left[ i({{\omega }_{A}}t+\beta \sin {{\omega }_{M}}t) \right],
\end{equation}
where $\beta $ is the maximum phase shift called the {\it modulation index}. We define the instantaneous frequency to be the time derivative of the phase of the field, which is in our case
\begin{equation}
\omega (t)=\frac{d}{dt}({{\omega }_{A}}t+\beta \sin {{\omega }_{M}}t)={{\omega }_{A}}+\Delta {{\omega }_{0}}\cos {{\omega }_{M}}t,
\end{equation}
where $\Delta {{\omega }_{0}}=\beta {{\omega }_{M}}$ is the maximum frequency shift called the frequency modulation {\it deviation}. Note that the phase modulation with a time dependence of $\beta \sin {{\omega }_{M}}t$ is equivalent to the frequency modulation with a time dependence of $\beta {{\omega }_{M}}\cos {{\omega }_{M}}t$. In the following, we assume $\beta \ll1$ (narrowband frequency modulation). The frequency-modulated incident laser field can then be decomposed into the three fields oscillating at ${{\omega }_{A}}$ (carrier) and ${{\omega }_{A}}\pm {{\omega }_{M}}$ (sidebands):
\begin{eqnarray}
\nonumber
{{E}_{I}} &\cong& {{E}_{0}}\exp (i{{\omega }_{A}}t)(1+i\beta \sin {{\omega }_{M}}t)\\
\nonumber
&=&{{E}_{0}}\exp (i{{\omega }_{A}}t)\left\{ 1+\frac{\beta }{2}\left[ \exp (i{{\omega }_{M}}t)-\exp (-i{{\omega }_{M}}t) \right] \right\}.\\
&&
\end{eqnarray}
Using Eq. (\ref{dipole2}) and the principle of superposition, the transmitted field ${{E}_{T}}$ is calculated to be 
\begin{eqnarray}
\nonumber
{{E}_{T}} & = & {{E}_{0}}\exp (i{{\omega }_{A}}t) \left[ \left( 1-\frac{\alpha l}{2} \right) \right. \\ 
\nonumber
&& +\frac{\beta }{2}\left( 1-\frac{\alpha l}{2}\frac{1-i{{x}_{M}}}{1+x_{M}^{2}} \right) \exp (i{{\omega }_{M}}t) \\
\nonumber
&& \left. -\frac{\beta }{2}\left( 1-\frac{\alpha l}{2}\frac{1+i{{x}_{M}}}{1+x_{M}^{2}} \right) \exp (-i{{\omega }_{M}}t)  \right] \\ 
\nonumber
&=&  {{E}_{0}}\exp (i{{\omega }_{A}}t)\left[ \left( 1-\frac{\alpha l}{2} \right)+i\beta \sin {{\omega }_{M}}t \right. \\
&&\left.+i\beta \frac{\alpha l}{2}\frac{1}{1+x_{M}^{2}}(x\cos {{\omega }_{M}}t-\sin {{\omega }_{M}}t) \right], 
\end{eqnarray}
where ${{x}_{M}}={{\omega }_{M}}/(\Gamma /2)$ is a normalized modulation frequency. The complex transmittance is then given by 
\begin{eqnarray}
\nonumber
\frac{{{E}_{T}}}{{{E}_{I}}}&=&\frac{\left( 1-\frac{\alpha l}{2} \right)+i\beta \left( \sin {{\omega }_{M}}t+\frac{\alpha l}{2}\frac{{{x}_{M}}\cos {{\omega }_{M}}t-\sin {{\omega }_{M}}t}{1+x_{M}^{2}} \right)}{1+i\beta \sin {{\omega }_{M}}t}\\
\nonumber
&\cong&\left( 1-\frac{\alpha l}{2} \right)+i\frac{\Delta {{\omega }_{0}}}{(\Gamma /2)}\frac{\alpha l}{2}\frac{x_{M}^{{}}\sin {{\omega }_{M}}t+\cos {{\omega }_{M}}t}{1+x_{M}^{2}},\\
\end{eqnarray}
where we neglect the terms proportional to ${{\beta }^{2}}$ and rewrite $\beta {{x}_{M}}=\beta {{\omega }_{M}}/(\Gamma /2)$ with the frequency modulation deviation $\Delta {{\omega }_{0}}=\beta {{\omega }_{M}}$ to elucidate the characteristics of the frequency discriminator. In polarization spectroscopy using a balanced polarimeter, the phase shift difference $\Delta \phi $ between the $\sigma^{+}$ and $\sigma^{-}$ components of the transmitted field is converted to a photocurrent signal $\Delta I$ with a relation $\Delta I=2{{I}_{T}}\Delta \theta ={{I}_{T}} \Delta \phi $, where ${{I}_{T}}$ is the photocurrent signal corresponding to the total transmitted beam power and $\Delta \theta =\Delta \phi /2$ is the rotation angle of the polarization axis. If the frequency modulation deviation $\Delta {{\omega }_{0}}$ is much smaller than the atomic natural width $\Gamma $ (namely, if the phase shift difference $\Delta \phi $ is much less than 1 radian), the photocurrent signal can be approximately expressed as 
\begin{eqnarray}
\nonumber
\Delta I&=&{{I}_{T}} \tan^{-1}\frac{\operatorname{Im}[{{E}_{T}}/{{E}_{I}}]}{\operatorname{Re}[{{E}_{T}}/{{E}_{I}}]}\cong {{I}_{T}}\frac{\operatorname{Im}[{{E}_{T}}/{{E}_{I}}]}{\operatorname{Re}[{{E}_{T}}/{{E}_{I}}]} \\
&=&\Delta {{I}_{\max }}\frac{x_{M}^{{}}\sin {{\omega }_{M}}t+\cos {{\omega }_{M}}t}{1+x_{M}^{2}},
\end{eqnarray}
where $\Delta {{I}_{\max }}={{I}_{T}}\alpha l\Delta {{\omega }_{0}}/(1-\alpha l/2)\Gamma$ is the maximum photocurrent signal amplitude obtained in the limit of ${{x}_{M}}\to 0$. If we regard the frequency modulation of $\Delta {{\omega }_{0}}\cos {{\omega }_{M}}t$ as an input and the resultant photocurrent signal $\Delta I$ as an output, the transfer function of this frequency discriminator is given by
\begin{equation}
\label{transfer2}
G({{\omega }_{M}})=\frac{{{G}_{0}}}{1+i\left( \frac{{{\omega }_{M}}}{\Gamma /2} \right)},
\end{equation}
where ${{G}_{0}}={{I}_{T}}\alpha l/(1-\alpha l/2)\Gamma$. This transfer function has the same form as that for a RC low-pass filter (a first-order Butterworth filter), which is characterized by a single-poll (6 dB/octave) roll-off and a 90${}^\circ$ phase lag above the cutoff frequency (see Fig. \ref{figA1}). Note that the 3-dB cutoff frequency of a discriminator based on polarization spectroscopy is $\Gamma /2$ rather than $\Gamma$. The transfer function of a PDH frequency discriminator has also the same form as that for a RC low-pass filter whose 3-dB cutoff frequency is $\Delta \omega_{C}/2$ (the HWHM width of the cavity resonance).
\begin{figure}[tttt]
\begin{center}
\includegraphics[width=85mm]{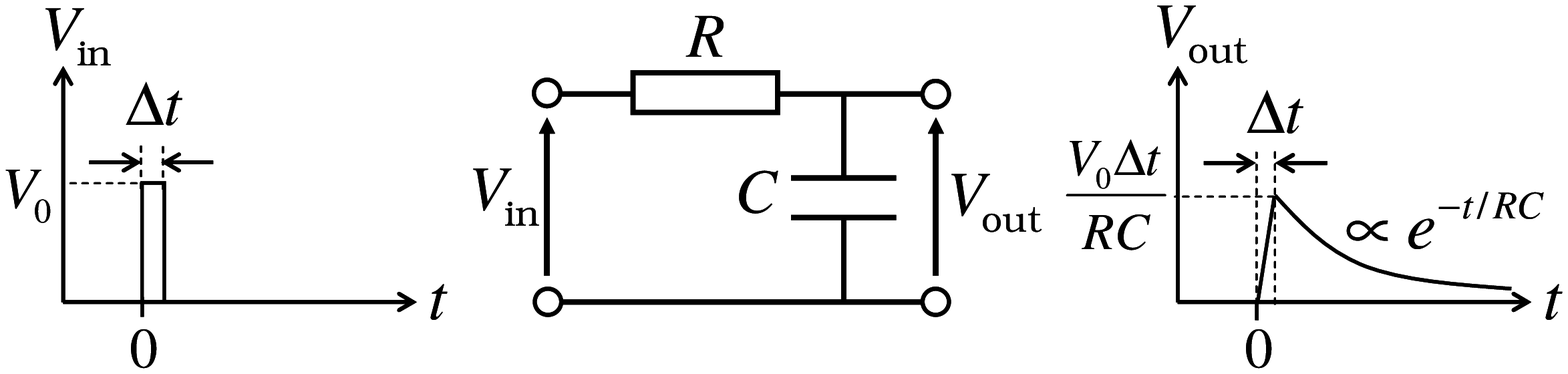}
\end{center}
\caption{Response of the RC low-pass filter to a rectangular voltage pulse with height $V_{0}$ and duration $\Delta t$.}
\label{figA2}
\end{figure}

The prompt response of a  polarization-spectroscopy-based (or PDH) frequency discriminator to a sudden phase jump in the incident laser field can be intuitively explained using the analogy of a RC low-pass filter composed of a resistance $R$ and a capacitance $C$ in series (Fig. \ref{figA2}). A phase jump $\Delta \varphi$ occuring in a short time $\Delta t\ll 1/\Gamma$ can be expressed as a rectangle-shaped time variation of the instantaneous frequency with height (frequency deviation) $\Delta \omega = \Delta \varphi / \Delta t$ and duration $\Delta t$. If we input a rectangular voltage pulse with height $V_{0}=\Delta \omega$ and duration $\Delta t$ to a RC low-pass filter whose cutoff frequency is $1/RC=\Gamma/2$, the resultant output is a linear voltage rise (charging of the capacitor) up to $V_{0} \Delta t /RC=\Delta \varphi\Gamma/2 $ during $\Delta t$, followed by an exponential decay with a time constant of $\tau=RC=2/\Gamma$. Recalling that the transfer function of this RC low-pass filter is exactly the same as Eq.(\ref{transfer2}) except for the gain $G_{0}$ in the passband, we can realize that the response of the discriminator to a sudden phase jump is a prompt rise of the photocurrent signal up to $\Delta I=G_{0}\Delta \varphi\Gamma/2$, which is proportional to the phase jump $\Delta \varphi$. This prompt response of the discriminator to a phase jump allows us to detect laser phase fluctuations beyond its cutoff frequency and effectively reduce the laser linewidth.

%\bibliography{ref}

%merlin.mbs apsrev4-1.bst 2010-07-25 4.21a (PWD, AO, DPC) hacked
%Control: key (0)
%Control: author (72) initials jnrlst
%Control: editor formatted (1) identically to author
%Control: production of article title (-1) disabled
%Control: page (0) single
%Control: year (1) truncated
%Control: production of eprint (0) enabled
%

\end{document}